\documentstyle[aps,prl,amssymb,latexsym,epsfig]{revtex}

\newcommand{\beq}{\begin{equation}}
\newcommand{\eeq}{\end{equation}}
\newcommand{\bqa}{\begin{eqnarray}}
\newcommand{\eqa}{\end{eqnarray}}
\newcommand{\fr}{\frac}

\begin{document}

\draft

\wideabs{
\title{Black hole formation from massive scalar field collapse in the Einstein-de Sitter universe}
\author{S\' ergio M. C. V. Gon\c calves}
\address{Theoretical Astrophysics, California Institute of Technology, Pasadena, California 91125}
\date{\today}
%\preprint{GRP-???}
\maketitle
\begin{abstract}
We study the spherically symmetric collapse of a real, minimally coupled, massive scalar field in an asymptotically Einstein-de Sitter spacetime background. By means of an eikonal approximation for the field and metric functions, we obtain a simple analytical criterion---involving the physical size and mass scales (the field's inverse Compton wavelength and the spacetime gravitational mass) of the initial matter configuration---for generic (non-time-symmetric) initial data to collapse to a black hole. This analytical condition can then be used to place constraints on the initial primordial black hole spectrum, by considering spherical density perturbations that re-entered the horizon during an early matter-dominated phase that immediately followed inflation.
\end{abstract}
}

\narrowtext

\section{INTRODUCTION}

Until now no fundamental elementary spin-0 particle has been detected in accelerator experiments, even though its existence is predicted by the so far highly successful $\mbox{SU}(3)\otimes\mbox{SU}(2)\otimes\mbox{U}(1)$ Standard Model \cite{donoghue&golowich&holsten92}. Despite this lack of knowledge, most models of early universe astroparticle physics include at least one such field, and their remarkable agreement with observational data (see e.g. \cite{guth00} for a recent review) makes pertinent the question of whether primordial density fluctuations in the scalar field energy density distribution could have collapsed to form black holes.

In particular, several models of inflation predict the existence of a period dominated by the energy of a real massive scalar field, just after the end of the inflationary epoch \cite{hawking&moss82,linde82,albrecht&steinhardt82}. During this period, the field can behave like dust and density inhomogeneities can undergo non-linear growth, which may lead to the formation of primordial black holes \cite{khlopov&malomed&zeldovich85}. It is during this early matter dominated phase that black hole formation---via gravitational collapse---is the most abundant, and we should therefore expect important constraints on inflationary models from the overproduction of primordial black holes during this epoch.

In this paper, we address one aspect of this problem by obtaining analytical conditions for black hole formation during the early matter dominated phase; the constraining of the primordial black hole energy density spectrum and its implications for inflationary scenarios will be discussed in detail elsewhere \cite{goncalves00}. 

Our model consists of a real, spherically symmetric massive scalar field, minimally coupled to gravity, evolving in an asymptotically Einstein-de Sitter (EdeS) spacetime background. The Einstein equations for gravity coupled to the massive scalar field are solved analytically, using an asymptotic expansion for the field (and metric functions), in terms of the field's Compton wavelength, $\mu^{-1}$, to explore the large mass limit. In such a limit, the scalar field behaves like general inhomogeneous dust \cite{goncalves&moss97}, described by the Tolman-Bondi metrics \cite{lemaitre33,tolman34,datt38,bondi48}. From generic initial data, spherical dust collapse always proceeds to a black hole\footnote{Central naked singularities can form from regular initial data, but they are {\em not} generic: only one non-spacelike geodesic can escape from the singularity, which, although strong, is also massless \cite{dwivedi&joshi92}.} and thus, by imposing the condition that the eikonal approximation holds, we obtain a sufficient criterion for black hole formation from massive scalar field collapse.

This paper is organized as follows: Section II derives the field equations for the massive Einstein-Klein-Gordon system, which are then analytically solved using an asymptotic series (in $\mu^{-1}$) for the field and metric functions. The large mass limit of the eikonal approximation is the Tolman-Bondi family of metrics, described in Sec. III. In Sec. IV, the gravitational collapse of spherical density perturbations in the EdeS universe---a particular case of the general Tolman-Bondi spacetimes---is analyzed in terms of initial data. In Sec. V, analytical conditions for the initial data to collapse to a black hole are obtained by enforcing the validity of the WKB approximation to second order in $\mu^{-1}$. Section VI discusses how this condition for black hole formation might affect the constraining of the initial (at the early matter dominated phase) primordial black hole spectrum, and thus the inflationary models responsible for generating the initial density fluctuation spectrum. Section VII concludes with a summary and a brief discussion on avenues for future work.

Geometrized units, in which $G=c=1$, are used throughout.

\section{THE MASSIVE EINSTEIN-KLEIN-GORDON SYSTEM}

We consider a general spherically symmetric metric, written here in normal Gaussian coordinates $\{t,r,\theta,\varphi\}$:
\beq
ds^{2}=-dt^{2}+e^{-2\Lambda}dr^{2}+R^{2}d\Omega^{2}, \label{ssmetric}
\eeq
where $\Lambda=\Lambda(t,r)$, $R=R(t,r)$, and $d\Omega^{2}=d\theta^{2}+\sin^{2}\theta d\varphi^{2}$ is the canonical metric of the unit 2-sphere.

The independent non-vanishing Einstein tensor components are
\bqa
G_{tt}&=&R^{-2}[-Re^{2\Lambda}(2R'\Lambda'+2R''+R^{-1}R'^{2}) \nonumber \\
&&-2\dot{R}\dot{\Lambda}R+1+\dot{R}^{2}], \\
G_{rt}&=&-2R^{-1}(\dot{R}'+R'\dot{\Lambda}), \\
G_{rr}&=&-R^{-2}\left[e^{-2\Lambda}(2\ddot{R}R+\dot{R}^{2}+1)-R'^{2}\right], \\
G_{\theta\theta}&=&\sin^{-2}\theta\,G_{\varphi\varphi}=R[\dot{R}\dot{\Lambda}+R'\Lambda'e^{2\Lambda} \nonumber \\
&&+R''e^{2\Lambda}-\ddot{R}+\ddot{\Lambda}R-\dot{\Lambda}^{2}R],
\eqa
where the overdot and prime denote partial differentiation with respect to $t$ and $r$, respectively.

For the matter content, we consider a real minimally coupled scalar field $\phi$ of mass $\mu$, governed by the Klein-Gordon equation:
\beq
(\Box-\mu^{2})\phi=0. \label{kgeq}
\eeq
With the spherically symmetric metric (\ref{ssmetric}) we have
\bqa
\ddot{\phi}-e^{2\Lambda}\phi''+\mu^{2}\phi+(\dot{\Lambda}+2R^{-1}\dot{R})\dot{\phi} \nonumber \\
-e^{2\Lambda}(\Lambda'-2R^{-1}R')\phi'=0.
\eqa
The stress-energy tensor of the scalar field is given by
\beq
T_{ab}=\nabla_{a}\phi\nabla_{b}\phi-\fr{1}{2}g_{ab}(\nabla_{c}\phi\nabla^{c}\phi+\mu^{2}\phi^{2}),
\eeq
with the independent non-vanishing components
\bqa
T_{tt}&=& \fr{1}{2}\dot{\phi}^{2}+\fr{1}{2}e^{2\Lambda}\phi'^{2}+\fr{1}{2}\mu^{2}\phi^{2}, \\
T_{rt}&=& \dot{\phi}\phi', \\
T_{r}^{r}&=& \fr{1}{2}\dot{\phi}^{2}+\fr{1}{2}e^{2\Lambda}\phi'^{2}-\fr{1}{2}\mu^{2}\phi^{2}, \\
T_{\theta\theta}&=&\sin^{-2}\theta\,T_{\varphi\varphi}=\fr{1}{2}R^{2}(\dot{\phi}^{2}-e^{2\Lambda}\phi'^{2}-\mu^{2}\phi^{2}).
\eqa

By defining the auxiliary functions
\bqa
k(t,r)&\equiv&1-e^{2\Lambda}, \label{kappa} \\
m(t,r)&\equiv&\fr{1}{2}R(\dot{R}^{2}+k), \label{mass}
\eqa
Einstein's equations can be recast in terms of the first derivatives of these two functions:
\bqa
k'&=&-8\pi RR'(T_{tt}+T^{r}_{r})-2R'(\ddot{R}+\dot{\Lambda}\dot{R}), \\
\dot{k}&=&8\pi RR'T_{t}^{r}, \label{kdot} \\
m'&=&4\pi R^{2}R'T_{tt}-4\pi R^{2}\dot{R}T_{rt}, \label{mprime} \\
\dot{m}&=&4\pi R^{2}R'T_{t}^{r}-4\pi R^{2}\dot{R}T_{r}^{r}. \label{mdot}
\eqa
Since there are only three independent functions to be determined and four equations, only three of these are independent, with the remaining one being a constraint. Since the scalar wave Eq. (\ref{kgeq}) is implied by the Einstein equations (it follows from the contracted Bianchi identities), we shall take it together with Eqs. (\ref{kdot}) and (\ref{mdot}) as our complete set. We take Eq. (\ref{mprime}) as the constraint equation, since it provides a simple relation between the initial data and the initial mass profile.

To facilitate the resolution of the field equations, we introduce a WKB approximation for the field in the large mass ($\mu$) limit---when the Compton wavelength of the scalar field, $\mu^{-1}$, is much smaller than the radius $\lambda$ of the spherical region where the field is non-vanishing:
\beq
\lambda\mu\gg1.
\eeq
In such a limit, we expect wave-like solutions with slowly-varying (with respect to $t$) amplitude:
\beq
\phi(t,r)=\mu^{-1}\Phi(t,r)\cos\,\mu t.
\eeq
The stress-energy tensor components are then
\bqa
T_{tt}&=&\fr{1}{2}\Phi^{2}-\fr{1}{2}\mu^{-1}\Phi\dot{\Phi}\sin\,2\mu t+\fr{1}{4}(\dot{\Phi}^{2}+e^{4\Lambda}\Phi'^{2}) \nonumber \\
&&\times(1+\cos\,2\mu t), \\
T_{rt}&=&-\fr{1}{2}\mu^{-1}\Phi\Phi'\sin\,2\mu t+\fr{1}{2}\mu^{-2}\dot{\Phi}\Phi'(1+\cos\,2\mu t), \\
T_{r}^{r}&=&-\fr{1}{2}\Phi^{2}\cos\, 2\mu t-\fr{1}{2}\mu^{-1}\Phi\dot{\Phi}\sin\, 2\mu t \nonumber \\
&&+\fr{1}{4}(\dot{\Phi}^{2}+e^{4\Lambda}\Phi'^{2})(1+\cos\, 2\mu t), \\
T_{\theta\theta}&=&\fr{1}{2}R^{2}[-\Phi^{2}\cos\,2\mu t-\mu^{-1}\Phi\dot{\Phi}\sin\,2\mu t \nonumber \\
&&+\fr{1}{2}\mu^{-2}(\dot{\Phi}^{2}-e^{4\Lambda}\Phi'^{2})(1+\cos\,2\mu t).
\eqa
The form of the above equations suggests a trigonometric expansion for $\Phi(t,r)$ of the form
\beq
\Phi=\Phi_{0}+\sum_{m=1}^{\infty}\sum_{n=1}^{\infty} \mu^{-m}(\Phi^{\rm c}_{mn}\cos\, n\mu t+\Phi^{\rm s}_{mn}\sin\, n\mu t).
\eeq
The other metric functions are expanded analogously. This expansion in inverse powers of the mass leads to an asymptotic series \cite{lagrange1788,bogoliubovetal81} in $\mu^{-1}$, and hence it gives an exact solution in the infinite mass limit and approximate one otherwise. This method is usually referred to as the Lagrange method or the method of averaging \cite{verhulst91}.

Substitution of the expanded metric functions and their derivatives in Eqs. (\ref{kgeq}), (\ref{mass}), (\ref{kdot}) and (\ref{mdot}) fixes the coefficients of all the trigonometric terms at each order in $\mu^{-1}$. Up to $O(\mu^{-2})$ we find that the only non-vanishing terms are:
\bqa
\Phi(t,r)&=&\Phi_{0}(t,r)+O(\mu^{-3}), \\
m(t,r)&=&m_{0}(t,r)+\mu^{-1}m_{12}(t,r)\sin\,2\mu t \nonumber \\
&&+\mu^{-2}m_{22}(t,r)\cos\,2\mu t+O(\mu^{-3}), \\
k(t,r)&=&k_{0}(t,r)+\mu^{-2}k_{22}(t,r)\cos\,2\mu t+O(\mu^{-3}), \\
R(t,r)&=&R_{0}(t,r)+\mu^{-2}R_{22}(t,r)\cos\,2\mu t+O(\mu^{-3}).
\eqa
Taking the first derivatives of the above equations and comparing the leading order terms to the right-hand-side of Eqs. (\ref{kgeq}), (\ref{mass}), (\ref{kdot}) and (\ref{mdot}), yields
\bqa
\dot{m}_{0}&=&0, \\
\dot{k}_{0}&=&0, \\
\dot{R}_{0}^{2}&=&2m_{0}R_{0}^{-1}-k_{0}.
\eqa
The leading order terms from Eq. (\ref{mprime}) give
\beq
m'_{0}=2\pi R_{0}^{2}R'_{0}\Phi^{2}_{0}.
\eeq
The first three equations restrict this class of metrics to the Tolman-Bondi family, which describes the collapse of general inhomogeneous dust. We have thence concluded that sufficiently---in a sense to be defined precisely below---massive scalar fields behave like a dust.

Spherical dust collapse is pressureless and thus always proceeds to a black hole. Therefore, by guaranteeing that the field behaves like dust until the complete formation of the event horizon, we can obtain a sufficient condition for black hole formation from massive scalar field collapse. This criterion is explicitly obtained by examining the next order terms in the WKB approximation. From Eqs. (\ref{mass}), (\ref{mdot}) and (\ref{kdot}), we have
\bqa
k_{22}(t,r)&=&2\pi R'_{0}(R_{0})^{-1}(1-k_{0})\Phi_{0}\Phi'_{0}, \\
m_{12}(t,r)&=&\pi R_{0}^{2}\dot{R}_{0}\Phi_{0}^{2}, \label{m12} \\
R_{22}(t,r)&=&-\fr{\pi}{2}R_{0}\Phi_{0}^{2}.
\eqa
These terms have the convenient feature of being simple algebraic functions of the known zeroth-order terms and their derivatives. The asymptotic expansion guarantees that the WKB approximation will remain valid whilst all the next order terms remain small compared to the leading order ones:
\beq
\mu^{-2}\fr{k_{22}}{k_{0}}\leq\fr{1}{2}\;\mbox{and}\;\mu^{2}\fr{|R_{22}|}{R_{0}}\leq\fr{1}{2}\;\mbox{and}\;\mu^{-1}\fr{m_{12}}{m_{0}}\leq\fr{1}{2}.
\eeq
The value $1/2$ was chosen arbitrarily (it has to be smaller than unity!) and changing this value amounts to a change in $\mu$. The WKB approximation will therefore breakdown when one of the above inequalities is no longer satisfied. The asymptotic expansion implies that the largest correction will come from the $m_{12}$ term, and we have verified this numerically. Thus, the validity of the WKB approximation is given by the condition [where Eqs. (\ref{mass}) and (\ref{m12}) were used]:
\beq
\dot{R}_{0}m'_{0}\leq\mu m_{0}R'_{0}, \label{wkb}
\eeq
which defines a region on the $t-r$ (hence $t-R$) plane, as shown in Fig.1. The WKB approximation will hold at all points in whose causal past the inequality is always satisfied. On the $t-R$ plane, it is the set of events {\em outside} the curve that saturates inequality (\ref{wkb}), whose past null cone is tangent to it. 

\begin{figure}
\begin{center}
%\leavevmode
\epsfxsize=15pc
\epsffile{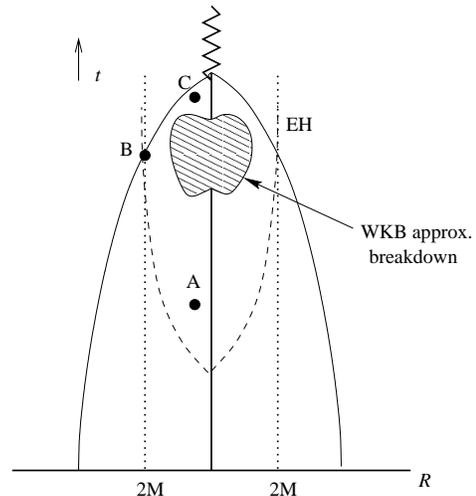}
\end{center}
\caption{Spherically symmetric dust collapse in $\{t,R\}$ coordinates. The solid line is the sphere's surface, and the dashed curve that asymptotes $R=2M$ is the event horizon. The patterned region inside the event horizon is where the WKB condition breaks down. The eikonal approximation holds for events $A$ and $B$, but not for event $C$: its past light cone intersects the region where the WKB condition breaks down. \label{fig1}}
\end{figure}

\section{THE TOLMAN-BONDI LIMIT}

From Eqs. (\ref{kappa}), (\ref{mass}), (\ref{kdot}), (\ref{mdot}), we have (hereafter dropping the `0'-indices for simplicity):
\bqa
ds^{2}&=&-dt^{2}+\fr{R'^{2}(t,r)}{1-k(r)}dr^{2}+R^{2}(t,r)d\Omega^{2}, \label{tbmetric} \\
\dot{R}^{2}(t,r)&=&\fr{2m(r)}{R(t,r)}-k(r), \label{tbdyn} \\
\Phi^{2}(t,r)&=&\fr{1}{2\pi}\fr{m'}{R^{2}R'}. \label{bigphi}
\eqa
This is the Tolman-Bondi class of solutions. Included in this class are the Schwarzschild metric [$m(r)=\mbox{const.}$], the Einstein-de Sitter universe [ $R(t,r)=a(t)r\propto t^{2/3}r$ and $k(r)=0$], and the closed Friedmann universe [$k(r)\propto r^{2}$]. The metric is written in comoving coordinates and the model describes the pressureless collapse of spherical dust shells. Each shell is labelled by the radial coordinate $r$ and has a surface area $4\pi R^{2}(t,r)$. In the context of Tolman-Bondi metrics, shell crossings---the overlapping of neighboring shells at finite proper radius---are defined by the locus of events for which $R>0$ and $R'=0$. It is clear from Eq. (\ref{bigphi}) that the energy density diverges at shell crossing singularities. Even though some curvature components may diverge at shell crossings \cite{yodzis&seifert&hagen73}, the spacetime is geodesically complete---analytical continuations of the metric can always be found (in a distributional sense) \cite{papapetrou67}---and thus they are not real physical singularities; rather, they merely signal the intersection of matter flow lines at a given spacelike surface, and thus lack of integrability of the field equations beyond that surface (it is worth pointing out that they also occur in inhomogeneous spherical Newtonian collapse).

Assuming $R'\neq0$, we can integrate Eq. (\ref{tbdyn}) parametrically to obtain
\bqa
t(\eta,r)&=&t_{0}(r)+\fr{m}{k^{3/2}}(\eta+\sin\eta), \label{time} \\
R(\eta,r)&=&\fr{2m}{k}\cos^{2}\fr{\eta}{2}, \label{aradius}
\eqa
where $t_{0}(r)$ is an arbitrary constant of integration (to be fixed by the initial data), and $0\leq\eta\leq\pi$. The physical initial data consists of a mass distribution $m(r)$ and a velocity profile $\dot{R}(0,r)$. There is gauge freedom for the scaling of $r$ and we shall fix it by requiring that it coincides with the initial area radius:
\beq
R(0,r)=r.
\eeq

\subsection{Time-symmetric initial data}

Considering time-symmetric initial data,
\beq
\dot{R}(0,r)=0,
\eeq
implies, from Eq. (\ref{tbdyn}), $k(r)=2m/r$ and $t_{0}(r)=0$. Equations (\ref{time}) and (\ref{aradius}) then simplify to
\bqa
t(\eta,r)&=&\left(\fr{r^{3}}{8m}\right)^{\fr{1}{2}}(\eta+\sin\eta), \\
R(\eta,r)&=&r\cos^{2}\fr{\eta}{2}.
\eqa
A shell with initial proper area $4\pi r^{2}$ will thus collapse to vanishing area radius in a (comoving) time
\beq
t_{\rm coll}(r)=\pi\sqrt{\fr{r^{3}}{8m}}.
\eeq
For inhomogeneous mass distributions ($m\neq\mbox{const.}\times r^{3}$), different shells become singular at different times; in the homogeneous case, all the shells collapse to zero area radius at the same time \cite{newtoniantoo}. 

A particular solution is specified by an initial mass function $m(r)$, given uniquely by the initial density distribution:
\beq
m(r)=2\pi \int_{0}^{r} \Phi^{2}(0,\bar{r})\bar{r}^{2} d\bar{r} =4\pi \int_{0}^{r} T_{tt}(0,\bar{r})\bar{r}^{2} d \bar{r}. \label{mass3}
\eeq
If the mass function $m(r)$ approaches a constant value $M$ when $r\rightarrow+\infty$, then $M$ is the ADM mass of the spacetime.

\section{THE EINSTEIN-DE SITTER UNIVERSE}

The EdeS universe is a particular member of the Friedmann-Robertson-Walker metrics\footnote{Which are a subclass of the Tolman-Bondi metrics [cf. Eq. (\ref{tbmetric})], with $R(t,r)=a(t)r$ and $k(r)=Kr^{2}$.},
\beq
ds^{2}=-dt^{2}+a^{2}(t)\left[\fr{dr^{2}}{1-Kr^{2}}+r^{2}d\Omega^{2}\right],
\eeq
where $K=-1,0,$ or $+1$, for a hyperbolic, flat or spherical spatial geometry, respectively. The Einstein equations are simply
\bqa
\dot{\rho}&=&-3(\rho+p)\fr{\dot{a}}{a}, \label{energycons} \\
\fr{\ddot{a}}{a}&=&-\fr{4\pi}{3}(\rho+3p), \label{raychaudhuri} \\
\left(\fr{\dot{a}}{a}\right)^{2}&=&\fr{8\pi}{3}\rho-\fr{K}{a^{2}}, \label{friedmann}
\eqa
where $\rho$ is the proper energy density and $p$ the pressure. For pressureless matter distributions ($p=0$) and vanishing spatial curvature ($K=0$)---the EdeS universe---the solution to the Friedmann equation (\ref{friedmann}) is
\beq
a(t)=a_{0}t^{\fr{2}{3}},
\eeq
where $a_{0}$ is an integration constant. This model describes an open geometry (the $K=0$ spatial sections are diffeomorphic to $\Bbb{R}^{3}$) in the presence of a constant non-zero energy density distribution. Even though the EdeS spacetime is conformally flat, its causal structure is quite different from asymptotically flat geometries. In particular, and unlike Minkowski or Schwarzschild, past null infinity for EdeS is spacelike, and thus past particle horizons exist [which can be observer (i.e., $r$) dependent].

The EdeS spacetime is a good approximation to the large scale structure of the universe during a matter dominated phase, when the averaged (over space and time) energy density evolves adiabatically and pressures are vanishingly small, as, e.g., immediately after inflation \cite{hawking&moss82,linde82,albrecht&steinhardt82}. We shall therefore adopt such a metric to model the collapse of overdensity perturbations in the early matter dominated phase that followed inflation.

\begin{figure}
\begin{center}
%\leavevmode
\epsfxsize=15pc
\epsffile{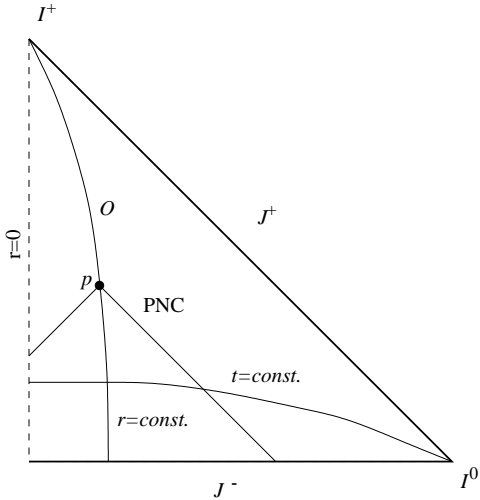}
\end{center}
\caption{Conformal diagram for the EdeS spacetime. The dashed vertical line on the left denotes the coordinate (central) singularity. PNC is the past null cone of a timelike observer $\cal{O}$ at event $p$. Worldlines starting from (spacelike) null infinity outside $p'\,$s PNC are causally disconnected from $\cal{O}$ at $p$. \label{fig2}}
\end{figure}

\subsection{Dust collapse in the Einstein-de Sitter universe}

Our model consists of a general Tolman-Bondi metric and a spherically symmetric matter distribution of finite proper radius $R(0,r)=r=\lambda$, with an overdensity $\zeta(r)$ with respect to the constant background energy density, $\rho_{\rm c}$:
\beq
T_{tt}(0,r)\equiv\rho(r)=\rho_{\rm c}[1+\xi(r)], \label{density}
\eeq
with $\xi(r)=\zeta(r)\Theta(r-\lambda)$, where $\zeta(r)$ is an arbitrary real-valued function and $\Theta$ is the unit Heaviside function. For $r\gg\lambda$, the metric asymptotically approaches the EdeS metric and the finite background energy density $\rho_{\rm c}$ implies the existence of a cosmological horizon with proper radius $H^{-1}=(8\pi\rho_{\rm c}/3)^{-1/2}$. Clearly, the existence of such a horizon (or, equivalently, expanding exterior geometry) precludes the use of time-symmetric initial data. 

Let us then define the generalized Hubble parameter by
\beq
H(t,r)\equiv\fr{\dot{R}}{R}.
\eeq
Then, from Eqs. (\ref{tbdyn}), (\ref{time}), (\ref{aradius}), we have
\bqa
H_{0}^{2}(r)&=&\fr{2m}{r^{3}}\sin^{2}\fr{\eta_{0}}{2}, \label{h0} \\
k(r)&=&\fr{2m}{r}\cos^{2}\fr{\eta_{0}}{2}=\fr{2m}{r}\left(1-\fr{H_{0}^{2}r^{3}}{2m}\right), \label{kay} \\
t_{0}(r)&=&-\fr{m}{k^{3/2}}[\eta_{0}(r)+\sin\eta_{0}(r)], \label{t0}
\eqa
where $H_{0}(r)\equiv H(0,r)=\dot{R}(0,r)/r$. Initial data consists of a mass profile $m(r)$ and an initial Hubble parameter $H_{0}(r)$. We shall consider here the case of an EdeS universe with initial expansion rate
\beq
H_{0}^{2}=\fr{8\pi}{3}\rho_{\rm c}.
\eeq
From Eqs. (\ref{mass3}) and (\ref{density}), the mass function is then
\bqa
m(r)&=&4\pi \int_{0}^{r} T_{tt}(0,x)x^{2}dx=\fr{4\pi}{3}\rho_{\rm c}r^{3}(1+\bar{\xi}) \nonumber \\
&=&M\left(\fr{r}{\lambda}\right)^{3}, \label{mass2}
\eqa
where $\bar{\xi}$ is the volume average of $\xi(r)$ and $M$ is the mass inside a sphere of proper area $4\pi R^{2}(t,\lambda)$.

\begin{figure}
\begin{center}
%\leavevmode
\epsfxsize=20pc
\epsffile{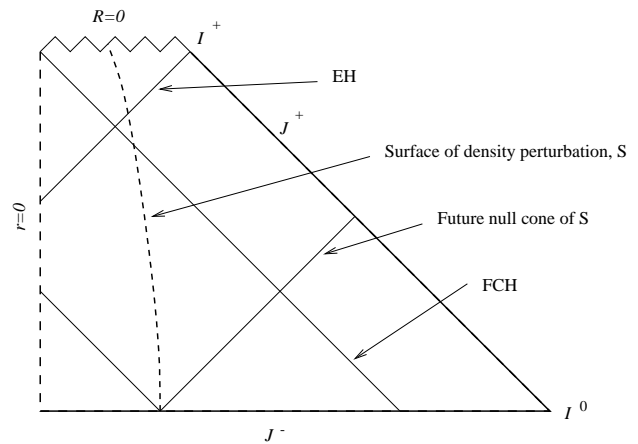}
\end{center}
\caption{Conformal diagram for spherical dust collapse in the EdeS universe. As the overdensity collapses (dashed curve), it first crosses the future cosmological horizon (FCH) of an $r=0$ observer, and, at a later time, the black hole event horizon, eventually ending up at the $R=0$ central singularity. Events outside the FCH will never be seen by the $r=0$ observer. Events outside the future null cone of S (the overdensity's spacelike 2-surface) will never be influenced by the overdensity perturbation.\label{fig3}}
\end{figure}

\section{CONDITIONS FOR BLACK HOLE FORMATION}

We want to impose the condition that the stationary phase approximation holds [cf. Eq. (\ref{wkb})]:
\beq
\dot{R}m'<\mu R'm, \label{wkb2}
\eeq
where
\bqa
R'&=&R\left(\fr{m'}{m}-\fr{k'}{k}\right)+\tan\left[t'_{0}\sqrt{k}+\fr{m}{k}\gamma(\eta+\sin\eta)\right], \\
\dot{R}&=&-\sqrt{k}\tan\fr{\eta}{2}, \\
t'_{0}&=&-\gamma t_{0}-\fr{m}{k^{3/2}}\cot\fr{\eta}{2}\cos^{2}\fr{\eta}{2}\left(\gamma+\fr{k'}{2k}-\fr{1}{r}\right), \label{tzerodot} \\
\gamma(r)&\equiv&\fr{m'}{m}-\fr{3}{2}\fr{k'}{k}. \label{gamma}
\eqa
For simplicity, we consider the energy density distribution inside the overdense region to be constant and hence, from Eqs. (\ref{mass2}), (\ref{tzerodot}) and (\ref{gamma}), we have $\gamma(r)=t'_{0}(r)=0$. The relevant WKB condition (\ref{wkb2}) becomes then
\beq
3p^{3}\sqrt{\fr{2M}{\lambda^{3}}}<\mu\cos^{2}\fr{\eta}{2}\cot\fr{\eta}{2}, \label{wkb3}
\eeq
where
\beq
p(\bar{\xi})\equiv\left(1+\bar{\xi}\right)^{-\fr{1}{2}} \in [0,1)
\eeq
is a monotonically increasing function of the magnitude of the volume-averaged overdensity, satisfying $p(0)=0$ and $\lim_{\bar{\xi}\rightarrow+\infty} p=1$. 

Outside the overdensity region, for $r>\lambda$, the Tolman-Bondi solution is exact: the EdeS metric is a particular case, trivially obtained from the general Tolman-Bondi form by setting $R(t,r)=(t/t_{\rm i})^{2/3}r$ and $k(r)=0$, where $t_{\rm i}$ is the initial slice. Therefore it suffices to enforce the above inequality (\ref{wkb3}) for $0<r\leq\lambda$, to ensure that the WKB approximation will hold for all $r>0$. If all the shells with $0<r\leq\lambda$ cross the event horizon {\em before} the WKB breaks down, then we can conclude that a black hole has formed.

In order to find the loci of events defined by the WKB condition and place them wholly inside the event horizon (EH), we need to (i) find a suitable parameterization for the curve that bounds the region defined by Eq. (\ref{wkb3}), and (ii) solve for the EH with the same parametrization for the null generators.

The natural such parameterization for the curve that saturates inequality (\ref{wkb3}) is $\eta_{*}(r)$ [ since $\eta$ and $p$ (through $\bar{\xi}$) depend solely on $r$], given by:
\beq
3p^{3}\sqrt{\fr{2M}{\lambda^{3}}}=\mu\cos^{2}\fr{\eta_{*}}{2}\cot\fr{\eta_{*}}{2}. \label{wkb4}
\eeq

From Eqs. (\ref{tbmetric}) and (\ref{time}), we can parameterize the null generators by $\eta_{\rm EH}$, obtained by integration of the outgoing radial null geodesics equation:
\beq
\fr{d\eta_{\rm EH}}{dr}=\left\{R'(k^{-1}-1)^{-\fr{1}{2}}-\left[t'_{0}\sqrt{k}+\fr{m}{k}\gamma(\eta+\sin\eta)\right]\right\}R^{-1}.
\eeq
For $r\leq\lambda$, this simplifies to
\beq
\fr{d\eta_{\rm EH}}{dr}=\left(\fr{\lambda^{3}}{2Mp^{2}}-r^{2}\right)^{-\fr{1}{2}},
\eeq
which is trivially integrated to give
\beq
\eta_{\rm EH}(r\leq\lambda)=\eta_{\rm EH}(0)+\sin^{-1}p\sqrt{\fr{2M}{\lambda}}\fr{r}{\lambda}.
\eeq
Outside the overdense region, $\gamma\neq0$ and $t'_{0}\neq0$, which makes it impossible to integrate the radial null geodesics equation analytically. Since the exterior geometry is not asymptotically flat, we cannot use Birkhoff's theorem to place the EH at $R=2M$. We can, however, solve for the apparent horizon (AH)---the outer boundary of a closed spacelike surface whose normal null geodesics have vanishing divergence---at any given $t=\mbox{const.}$ hypersurface. In spherical symmetry, the equation for the AH is simply the requirement that a surface $R(t,r)=\mbox{const.}$ becomes null:
\beq
R^{,a}R_{,a}=0.
\eeq
From Eqs. (\ref{tbmetric}) and (\ref{tbdyn}), we have then
\beq
R(t_{\rm AH},r)=2m(r),
\eeq
which, via Eq. (\ref{time}), defines a curve $\eta_{\rm AH}(r)$, given by
\beq
\eta_{\rm AH}(r)=2\cos^{-1}\fr{r}{\lambda}\sqrt{\fr{2M}{\lambda}}p.
\eeq
Since the AH must be contained or coincide with the (spacelike section of the) EH, we must have
$\eta_{\rm AH}(\lambda)\geq\eta_{\rm EH}(\lambda)$, which gives an upper bound for $\eta_{\rm EH}(0)\equiv\eta_{0}$:
\beq
\eta_{0}\leq2\cos^{-1}\sqrt{\fr{2M}{\lambda}}p-\sin^{-1}\sqrt{\fr{2M}{\lambda}}p.
\eeq
This upper bound on $\eta_{0}$ guarantees that the approximate EH thus estimated coincides or lies inside the actual EH; hence, if the WKB approximation holds inside the former, it will obviously hold inside the latter. 

We now need to guarantee that the breakdown domain of the WKB approximation is placed entirely inside the EH. From Eq. (\ref{wkb4}), $\eta_{*}$ is implicitly defined by an expression of the form $f[\eta_{*}]=\fr{1}{3}p^{-3}\sqrt{\fr{\lambda^{3}}{2M}}$, where $f$ is a monotonically increasing functional of $\eta_{*}$; therefore, $\eta_{*}>\eta_{\rm EH}(\lambda)\,\Rightarrow\, f[\eta_{*}]>f[\eta_{\rm EH}(\lambda)]$. This enables a sufficient condition for the WKB approximation to hold (inside the true EH), by imposing a lower bound on $\eta_{*}$:
\beq
\eta_{*}>\eta_{\rm EH}(\lambda)\geq2\cos^{-1}\sqrt{\fr{2M}{\lambda}}p. 
\eeq
From the above equation, together with Eq. (\ref{wkb3}), we finally obtain
\beq
\mu M>\fr{3}{2}\sqrt{1-\fr{2M}{\lambda}p^{2}}.
\eeq
Matter configurations with parameters satisfying the above inequality will collapse to form a black hole. This is a sufficient, but not necessary, criterion, due to both the adiabatic ansatz $(\mu\lambda\gg1)$ and the approximation for the EH. There are two limits that should be pointed out:

\subsubsection*{A.1. $p\rightarrow1$} 
In the limit of very large overdensities, the inequality approaches the one for the asymptotically flat case \cite{goncalves&moss97}, as expected. The simple physical reason is that the ratio $\rho_{\rm c}/\rho_{\rm pert.}\ll1$, and therefore having a vanishing or finite background energy density becomes irrelevant; the overdense region does not ``feel'' the exterior geometry. This can be heuristically understood as follows: The typical cosmological expansion rate is $(\dot{R}/R)_{\rm exp}\sim\rho_{\rm c}^{-1/2}$, whereas the collapse rate is $(\dot{R}/R)_{\rm coll}\sim-[(1+\bar{\xi})\rho_{\rm c}]^{-1/2}$. Then, $\fr{\tau_{\rm exp}}{\tau_{\rm coll}}=\left|\left(\fr{\dot{R}}{R}\right)_{\rm coll}\right|/\left(\fr{\dot{R}}{R}\right)_{\rm exp}\sim(1+\bar{\xi})^{1/2}\gg1$: the expansion that the overdensity would have experienced during a (proper) time $t_{\rm coll}$, had it not collapsed during that time, is negligible compared to its initial physical radius.

\subsubsection*{A.2. $p\rightarrow0$} 
When the overdensity vanishes, although the right-hand-side of the inequality approaches $3/2$, this result is meaningless because it was based on an estimate for the EH location that becomes totally unreliable: $\eta_{\rm EH}(0)=\pi$, when $p=0$. This is trivial, since $p=0$ corresponds to an unperturbed EdeS universe. For finite $p\ll1$, provided the spatial extent of the overdensity is sufficiently large, $\lambda>2Mp^{2}$, the inequality is still valid.

We can also look at this condition in terms of the cosmological horizon, $H_{0}^{-1}$. Noting that $2M=H_{0}^{2}\lambda^{3}(1-p^{2})^{-1}$ and recalling that $p\equiv(1+\bar{\xi})^{-1/2}$, we can rewrite the condition for black hole formation as
\beq
\mu M>\fr{3}{2}\sqrt{1-(H_{0}\lambda)^{2}\bar{\xi}}.
\eeq
The two relevant limits are now:

\subsubsection*{B.1. $H_{0}\lambda\ll1$} 
The length scale $\lambda$ is much smaller than the horizon radius $H_{0}^{-1}$, and $\bar{\xi}$ can be sufficiently large to recover the $p\rightarrow1$ limit and thus the asymptotically flat results. This limit also provides a useful self-consistency check for the validity of the WKB approximation. Let us take $H_{0}\lambda\ll1$ and $\bar{\xi}\sim O(1)$; then, the inequality becomes $H_{0}^{2}\lambda^{3}\mu\gtrsim\fr{3}{2}$. Since, by hypothesis, $H_{0}\lambda\ll1$, the necessary and sufficient condition to enforce the inequality is $\mu\lambda\gg1$, which is precisely the adiabatic condition on which the whole argument is based. Thus, we confirm that if the WKB approximation is valid initially, it will remain valid up to the formation of the EH.

\subsubsection*{B.2. $H_{0}\lambda\simeq1$} 
If $\bar{\xi}>(H_{0}\lambda)^{-2}$, then the right-hand-side of the inequality becomes imaginary and the approximation breaks down. Physically, this corresponds to an overdensity that becomes so large that a spacelike hypersurface would curve upon itself and form a disjoint spatial universe, which does not correspond to a black hole \cite{carr&hawking74}. Provided $\bar{\xi}<(H_{0}\lambda)^{-2}\lesssim1$, this limit is equivalent to the $p\ll1$ limit previously described, and thus still valid. From an astrophysical viewpoint, this is the most interesting limit, as it constrains the density perturbations at horizon crossing.

\section{CONSTRAINING THE PRIMORDIAL BLACK HOLE SPECTRUM}

Even though inflation is still a scenario and not yet a paradigm for cosmology, its basic predictions appear fairly robust in that not only they solve the (``naturalness'' and ``fine-tuning'') shortcomings of the ``old''standard Big Bang model, but they also present a viable physical mechanism for the primordial density fluctuations that seeded the subsequent structure formation.

As previously mentioned, several inflationary models predict the existence of a period dominated by the energy density of a real massive scalar field, just after the end of inflation, hereafter denoted by $t_{\rm end}$. During this stage, the field undergoes coherent oscillations, behaving effectively like a dust, from $t_{\rm end}$ until $t_{\rm RH}$ (``reheating''), when it rapidly decays into relativistic particles \cite{kolb&turner94}. Any density perturbations crossing the horizon during this dust-like phase will inevitably (there is no pressure) collapse to form black holes, provided they are sufficiently spherical, as aspherical growth precludes the gravitational collapse of the inhomogeneities \cite{khlopov&polnarev80}. Clearly, it is during this intermediate matter-dominated phase that black hole formation via gravitational collapse of (spherical) overdensities is the most abundant. We should therefore expect important constraints on inflationary scenarios (e.g., on the decay width of the scalar field, $\Gamma$) from the overproduction of primordial black holes (PBH) during this period.

The present process will only be relevant for black hole formation if the time interval, $\Gamma^{-1}$, after which the field decays into radiation is larger than the collapse time to the central singularity. We note that it is sufficient to require that the scalar field dominates just until the overdense region has collapsed through its Schwarzschild radius, since a subsequent decay into radiation (before vanishing proper area is reached) would not destroy the already formed black hole; furthermore, the proper time to reach the formation of black hole horizons (i.e., the proper time elapsed between horizon crossing and the emission of the null generators) is much larger than the time it would take for a black-hole-sized configuration to collapse to a singularity.

Since spherical symmetry is assumed throughout, the necessary and sufficient conditions for black hole formation from a massive scalar field configuration with parameters $\{\mu,M,\lambda\}$ are that black holes:
\bqa
&&\mbox{{\it have time} to form:}\; t_{\rm HC}+t_{\rm coll}<t_{\rm RH}, \label{timetoform} \\
&&\mbox{{\it will} form:}\; \mu M>\fr{3}{2}\sqrt{1-(H_{0}\lambda)^{2}\bar{\xi}}, \label{wkb5}
\eqa
where in the first inequality, the subscripts refer to ``horizon crossing'', ``collapse'' and ``reheating'', respectively. Because the synchronous gauge is being used, the three timescales can be trivially compared, thus justifying inequality (\ref{timetoform}). Since we are considering homogeneous density
distributions, the proper time for the collapse of the overdense region
will depend solely on its density. Setting $\eta=\pi$, from Eqs. (\ref{time}), (\ref{aradius}) and (\ref{h0})-(\ref{t0}), we
obtain
\bqa
t_{\rm
coll}&=&\fr{1}{2}H_{0}^{-1}\bar{\xi}^{-\fr{3}{2}}(1+\bar{\xi})(\pi-\sin^{-1}
\chi-\chi), \\
\chi&\equiv&2\sqrt{\bar{\xi}}(1+\bar{\xi})^{-1}.
\eqa
For large overdensities, $\bar{\xi}\gg1$, $t_{\rm coll}\sim\fr{1}{2}\pi
H_{0}^{-1}\bar{\xi}^{-\fr{3}{2}}$, whereas in the limit of small
perturbations, $\bar{\xi}\ll1$, the collapse time is
\beq
t_{\rm coll}\approx\fr{\pi}{2}H_{0}^{-1}\bar{\xi}^{-\fr{3}{2}}. \label{tc}
\eeq

In the EdeS universe, $H_{0}=\fr{2}{3}t^{-1}=\sqrt{8\pi\rho_{\rm c}/3}$.
At horizon crossing, $H\lambda=1$; thus
\beq
t_{\rm HC}=\fr{2}{3}\lambda=\fr{2}{3}H_{0}^{-1}.
\eeq
With Eq. (\ref{tc}), we can rewrite Eq. (\ref{timetoform}) as (dropping the bar for
simplicity)
\beq
\xi>\left(\fr{\pi}{2}\right)^{\fr{2}{3}}\left(H_{0}t_{\rm
RH}-\fr{2}{3}\right)^{-\fr{2}{3}}\equiv\xi_{\rm min}^{(1)}. \label{ximin1}
\eeq
That is, at a given slice $t=\fr{2}{3}H_{0}^{-1}$, there will be a minimum
fractional density perturbation at horizon crossing, below which the
scalar field will decay into radiation before the overdensity has time to
undergo complete gravitational collapse.

Let us now examine the WKB condition at horizon crossing. The mass of a
perturbation with fractional overdensity $\xi$, at horizon crossing, is
\bqa
M&=&\fr{4}{3}\pi\rho_{\rm
c}(1+\xi)\lambda^{3}=\fr{1}{2}H_{0}^{2}(1+\xi)\lambda^{3} \nonumber \\
&=&\fr{1}{2}H_{0}^{-1}(1+\xi)\simeq\fr{1}{2}H_{0}^{-1}.
\eqa
Equation (\ref{wkb5}) becomes then
\beq
\mu M>\fr{3}{2}\sqrt{1-\xi},
\eeq
which places another lower limit on $\xi$:
\beq
\xi>1-\fr{4}{9}\mu^{2}M^{2}\equiv\xi_{\rm min}^{(2)}. \label{ximin2}
\eeq

We have then two independent lower bounds on $\xi$, that tell us,
respectively, that black holes {\em can} and {\em will} form. If $\mu
M>\fr{3}{2}$, then the dust approximation remains reliable and Eq.
(\ref{ximin2}) is always satisfied. In this case, Eq. (\ref{ximin1})
becomes the relevant constraint, and the crucial quantity is $H_{0}t_{\rm
RH}$, which, although model dependent (through $t_{\rm RH}$, or,
equivalently, $\Gamma^{-1}$, for a given horizon crossing time), is
expected to be large, $H_{0}t_{\rm RH}>10^{6}$
\cite{kolb&turner94}.

If $\mu M<3/2$, then the dust approximation may become unreliable and the
fractional density perturbations are of order unity, therefore departing
from the linear regime and being heavily suppressed in the initial
perturbation spectrum, in which case the dust-like phase is irrelevant for
the production of PBH from the gravitational collapse
of massive scalar fields.

We shall then consider only configurations satisfying $\mu M>3/2$. Let us
first rewrite the density contrast in more familiar notation:
\beq
\xi=\fr{\rho}{\rho_{\rm c}}-1=\fr{\Delta\rho}{\rho_{\rm c}}\equiv\delta,
\eeq
where $\delta$ is to be evaluated at horizon crossing.

For the spectrum of density fluctuations crossing the horizon, we consider
a Gaussian probability distribution (as predicted by inflation
\cite{kolb&turner94}):
\beq
P(\delta)=\fr{1}{\sqrt{2\pi}\sigma}\exp\left(-\fr{\delta^{2}}{2\sigma^{2}}\right),
\eeq
where $\sigma$ is the root-mean-square amplitude of the density
fluctuations. Provided we restrict ourselves to linear perturbations, the
Fourier modes will evolve independently, thereby preserving Gaussianity.
(Non-Gaussian effects arise if $\delta\gg\sigma$, when
linearity is spoiled by mode-mode coupling, but their influence on the initial PBH mass function for a dust-dominated era has been shown to be negligible \cite{bullock&primack97}).

The quantity of interest in the constraining of PBH
production is the mass fraction of the universe that collapsed to a black
hole during the intermediate dust-like epoch:
\beq
\beta\equiv\fr{\rho_{\rm PBH}}{\rho_{\rm c}}=\int_{\xi_{\rm min}}^{1}
P(\delta)d\delta,
\eeq
where $\rho_{\rm PBH}$ is the initial (i.e., formed during that epoch)
energy density of primordial black holes, and $\rho_{\rm c}$ is the
background energy density. $\xi_{\rm min}$ is given by Eq. (\ref{ximin1})
and the upper integration limit is imposed by linearity. Expressed in this
way, $\beta$ is model dependent, since $\xi_{\rm min}$ depends on $H_{0}$
and $t_{\rm RH}$ (equivalently, $\Gamma^{-1}$).

$\beta$ can be observationally constrained via (i) the requirement that
the present density fraction of PBH's, $\Omega_{\rm PBH}<1$; and (ii)
limits from the Hawking radiation, from upper bounds on the PBH population at subsequent epochs, e.g., baryogenesis and  nucleosynthesis. These constraints on $\beta$
will lead to constraints on the cosmological parameters $\beta$ is
implicitly dependent on, namely the decay width of the scalar field and
the time for horizon crossing. Work in this direction is currently underway
and will be presented in detail elsewhere \cite{goncalves00}.

\section{CONCLUSIONS}

We have shown that there is an adiabatic regime ($\lambda\mu\gg1$), in which massive scalar fields in an asymptotically EdeS background behave like a dust and therefore inevitably collapse to form black holes. Enforcing the adiabacity condition to second order in $\mu^{-1}$ lead to a simple analytical criterion to be obeyed by the initial data to collapse to black holes. This criterion is valid for spherical density perturbations of {\em any} finite physical radius $\lambda$ ; in particular, they can equal the horizon size, $H\lambda=1$. Perturbations with spatial radius larger than the horizon radius are physically uninteresting because (i) the fractional density perturbation required to satisfy the WKB condition is very small, $\xi\ll(H\lambda)^{-2}$, and (ii) they are causally disconnected from the observable universe.

Immediately after inflation, there can exist a short period dominated by the energy density of a real massive scalar field \cite{hawking&moss82,linde82,albrecht&steinhardt82}. By considering spherical density perturbations that re-enter the horizon during this epoch, we obtained a lower limit on the fractional density perturbation (at horizon crossing) that would allow black holes to form before the field decays into relativistic particles. This lower limit can then be used to compute the initial mass function of PBH at the early matter-dominated phase. Constraints on the PBH mass spectrum from its subsequent evolution will ultimately lead to constraints on some free parameters of particle astrophysics models \cite{goncalves00}. 

\section*{ACKNOWLEDGMENTS}

I am grateful to Bernard Carr, Ian Moss, Kip Thorne and Michele Vallisneri for helpful discussions. This work was supported by F.C.T. (Portugal) Grant PRAXIS XXI-BPD-16301-98, and by NSF Grant AST-9731698.

\end{document}